\documentclass[aps,prb,preprint,showpacs,superscriptaddress]{revtex4}

\usepackage{epsf,graphicx}
\usepackage{amsmath}
\usepackage{nicefrac}
\usepackage{booktabs}

\usepackage{color}

\begin{document}


\title{Identification of coherent lattice modulations coupled to charge and orbital order in a manganite}


\author{A.~Caviezel}
\author{S.~O.~Mariager}
\affiliation{Swiss Light Source, Paul Scherrer Institut, 5232 Villigen PSI, Switzerland}
\author{S.~L.~Johnson}
\affiliation{Institute for Quantum Electronics, Physics Department, ETH Zurich, 8093 Zurich,
Switzerland}
\author{E.~M\"{o}hr-Vorobeva}
\altaffiliation{Present address: Department of Physics, Clarendon Laboratory, University of Oxford,
Parks Road, Oxford OX1 3PU, United Kingdom}
\author{S.~W.~Huang}
\author{G.~Ingold}
\author{U.~Staub}
\affiliation{Swiss Light Source, Paul Scherrer Institut, 5232 Villigen PSI, Switzerland}
\author{C.~J.~Milne}
\altaffiliation{Present address: Paul Scherrer Institut, CH-5232 Villigen PSI, Switzerland}
\affiliation{Laboratoire de Spectroscopie Ultrarapide, Ecole Polytechnique F\'ed\'erale de Lausanne, 1015 Lausanne, Switzerland}
\author{S.-W.~Cheong}
\affiliation{Rutgers Center for Emergent Materials and Department of Physics \& Astronomy, Rutgers University, Piscataway, NJ 08854, USA}
\author{P.~Beaud}
\email[Electronic address: ]{paul.beaud@psi.ch}
\affiliation{Swiss Light Source, Paul Scherrer Institut, 5232 Villigen PSI, Switzerland}


\date{\today}

\begin{abstract}
We apply grazing-incidence femtosecond x-ray diffraction to investigate the details of the atomic motion connected with a displacively excited coherent optical phonon. We concentrate on the low frequency phonon associated with the charge and orbital order in the mixed valence manganite La$_{0.25}$Pr$_{0.375}$Ca$_{0.375}$MnO$_3$ for $T \lesssim 210$ K. We measure the response of three superlattice reflections that feature different sensitivities to the motion of the unit cell constituents. The results support the assignment to a translational mode of the Mn$^{4+}$ atoms together with the oxygen atoms connecting adjacent Mn$^{4+}$ sites.
\end{abstract}


\maketitle

\section{\label{introduction}Introduction}
Correlated materials are known for their intricate balance of competing structural, magnetic as well as charge interactions and have been extensively studied over the past decades. Most studies focus on macroscopic properties in thermodynamic equilibrium as a function of external parameters such as temperature, chemical composition, magnetic or electric field. More recently time resolved experimental techniques have been applied to these systems with the goal to advance our understanding of the underlying correlations by investigating the interaction of the atomic, electronic and magnetic subsystems on their relevant time scales.\cite{Fiebig1998,Kimel2004,Polli2007,Matsubara2007,Foerst2011} In particular femtosecond x-ray or electron diffraction received considerable attention enabling direct access to the time resolved atomic and electronic structure.\cite{Sokolowski2003,Cavalleri2005,Eichberger2010,Mohr-Vorobeva2011,Johnson2012}

One of the strengths of diffraction techniques is the ability to fully determine the crystallographic unit cell by measuring a sufficient number of Bragg reflections. Consequently, this stimulated efforts to use time-resolved diffraction techniques to make movies of the atomic lattices in response to a selective fast perturbation of the system. This has been demonstrated in the picosecond regime applying Laue crystallography to a large biological molecule where a single diffraction image contains more than 3000 spots and thus holds a tremendous amount of information of the unit cell electron density.\cite{Schotte2003} For solid materials, however, the natural time scale of atomic motion is determined by the the period of optical phonons that typically lie in the range of 50 - 500 fs. Large bandwidth, so-called 'white' x-ray sources with sufficient flux as required for Laue crystallography are currently not available for femtosecond studies. Consequently, the complete reconstruction of the structural motion requires the measurement of numerous individual Bragg reflections obtained under identical excitation conditions. The feasibility to produce an atomic movie in the femtosecond time scale has been recently demonstrated in photo-excited tellurium.\cite{Johnson2009} Because the number of reflections required scales linearly with the number of atoms within the unit cell, unambiguous reconstruction becomes a challenge for more complex structures.

Doped manganite oxides of the general form $R_{1-x}A_x$MnO$_3$, which are transition metal oxides with a perovskite-type structure, exhibit extremely rich phase diagrams due to the subtle balance between charge, orbital, spin and structural degrees of freedom. Here, $R$ is a rare earth 3+ cation and $A$ is a 2+/1+ cation as the alkaline ions Ca, Na, Sr. In the doping region where these manganites exhibit colossal magnetoresistance it has been shown that sufficiently intense ultrashort optical pulses are able to induce an ultrafast metal-to-insulator transition in Pr$_{0.7}$Ca$_{0.3}$MnO$_3$.\cite{Fiebig1998,Polli2007} We have recently shown using time resolved x-ray diffraction that laser-induced melting of the charge and orbital order is accompanied by a symmetry change of the atomic lattice occurring on a non-thermal time scale in La$_{0.42}$Ca$_{0.58}$MnO$_3$ and La$_{0.25}$Pr$_{0.375}$Ca$_{0.375}$MnO$_3$.\cite{Beaud2009, Caviezel2012} This extremely fast structural transition is driven by the immediate release of the Jahn-Teller distortions of the Mn$^{3+}$O$_6$ octahedra that are intrinsically connected with the charge and orbital order (CO/OO).
\begin{figure*}[b]
\centering
\includegraphics[width=0.97\textwidth]{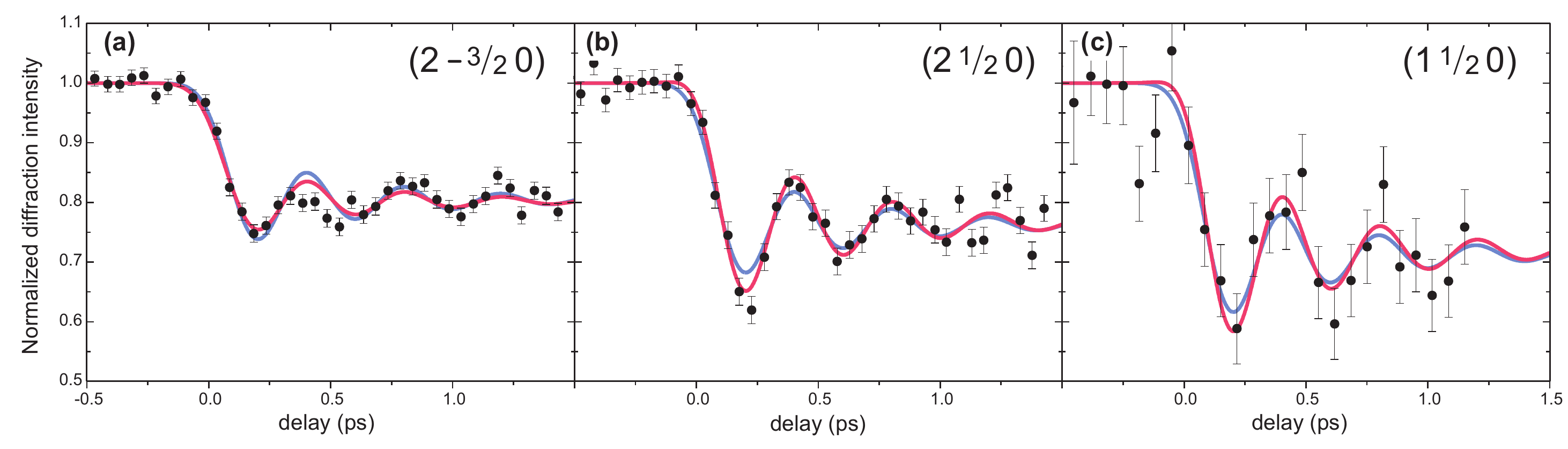}
\caption{Temporal evolution of the normalized diffracted x-ray intensity for the three superlattice reflections (2 -$\nicefrac{3}{2}$ 0),  ($2\:\nicefrac{1}{2}\:0$) and ($1\:\nicefrac{1}{2}\:0$) at an average excitation fluence $F_{\mathrm{avg}} = 1.6$ mJ/cm$^2$. The solid lines are fits obtained by a simple displacive excitation model (blue) and an extended two-step displacive excitation model (red), respectively.}\label{fig:TimeTraces}
\end{figure*}

Here, we use time resolved x-ray diffraction to identify the atomic motion connected with the low frequency coherent optical phonon of $A_g$ symmetry prominently observed when exciting manganites in the CO/OO phase.\cite{Lim2005,Beaud2009,Matsuzaki2009,Jang2010,Caviezel2012} Although it is not yet possible to create an atomic movie on such a complex system we will show that using a careful choice of reflections combined with structure factor calculations allow drawing conclusions about the underlying atomic motions.

\section{Experiment}
The single crystal of La$_{0.25}$Pr$_{0.375}$Ca$_{0.375}$MnO$_3$ (LPCMO) was grown in an optical floating-zone furnace \cite{Lee2002} and cut and polished to a (010) surface orientation labeled according to the room temperature unit cell notation in $Pbnm$ symmetry.\cite{Rodriguez2005} The time resolved x-ray diffraction experiments were performed at the hard x-ray FEMTO slicing facility at the Swiss Light Source.\cite{Beaud2007} In order to minimize fluorescence the experiments were performed with an x-ray energy of 5.0 keV which lies below the Mn $K$ and Pr $L$ edges. Details of the experimental grazing incidence setup are described elsewhere.\cite{Beaud2007,Caviezel2012} The sample was placed on a cryogenically cooled copper block thermally stabilized to a temperature of 150 K located on a hexapod sample manipulator in a vacuum chamber. The LPCMO crystal is excited via a weakly focused 100 fs laser pulse with $\lambda = 800$ nm and p-polarization. The pump fluence was set to $F= 3.9$ mJ/cm$^2$. A grazing incidence angle of 0.85$^{\circ}$ leads to an x-ray intensity attenuation length $z_{\mathrm{x}} = 90\pm10$ nm. Using the laser penetration depth $z_{\mathrm{L}} = 65\pm5$ nm the average excitation fluence lies well below the phase transition fluence $F_{\mathrm{th}}\approx$ 10 mJ/cm$^2$ at 150 K.\cite{Caviezel2012} The temporal resolution of 200$\pm$20 fs (FWHM) and the coincidence time $t_0$ between the optical pump and x-ray probe pulses were determined by measuring the ultrafast drop of a superlattice peak at high fluence.\cite{Beaud2009,Caviezel2012}

\section{Results and Discussion}
Figure \ref{fig:TimeTraces} shows the photoinduced response of the diffracted x-ray intensity for three different superlattice peaks (2 -$\nicefrac{3}{2}$ 0), ($2\:\nicefrac{1}{2}\:0$) and ($1\:\nicefrac{1}{2}\:0$). At 150 K LPCMO is an antiferromagnetic insulator exhibiting charge and orbital order. Because the structural distortions from the Jahn-Teller effect are relatively small, the resulting superlattice reflections are very weak yielding roughly 14 diffracted photons per second (ph/s) for the (2 -$\nicefrac{3}{2}$ 0), reflection, 9 ph/s for ($2\:\nicefrac{1}{2}\:0$) and 0.2 ph/s for ($1\:\nicefrac{1}{2}\:0$). Thus integration times up to 24 hours are needed for a sufficient signal to noise ratio. Nonetheless, as shown in Fig. \ref{fig:TimeTraces} a laser-induced phonon with a frequency of approximately 2.5 THz is clearly visible. For all three reflections we observe the typical signature of a coherent phonon excited via the displacive excitation mechanism: a fast change of diffraction intensity followed by coherent oscillations.\cite{Sokolowski2003}
\begin{figure*}[t]
\centering
\includegraphics[width=0.97\textwidth]{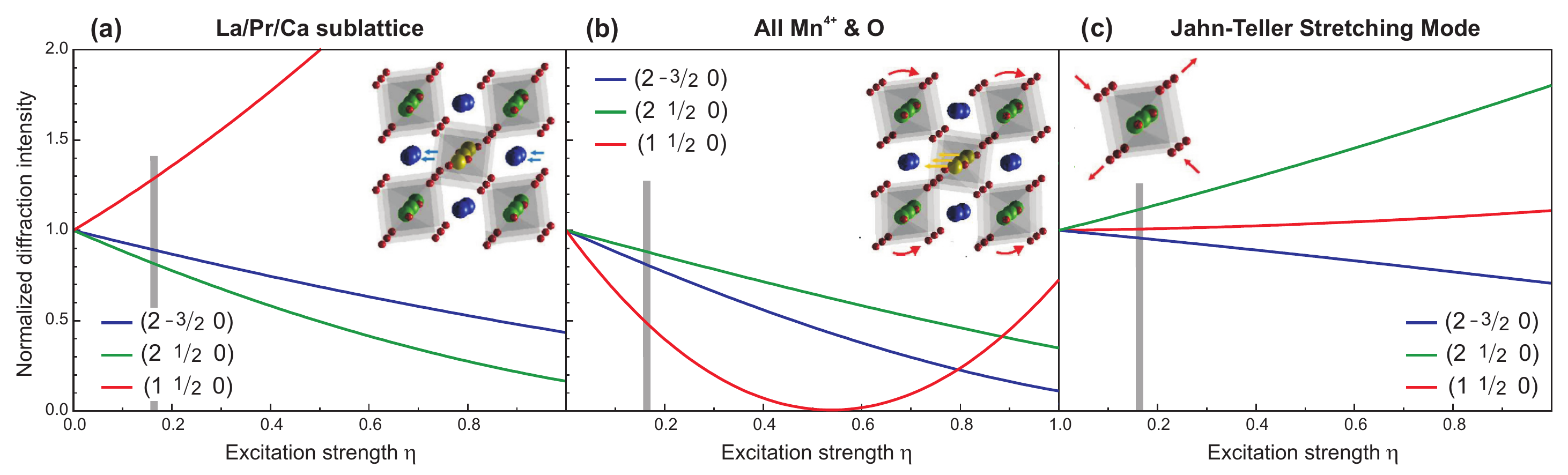}
\caption{Calculated normalized diffraction intensity and their visualization for three different atomic displacements: (a) motion of the La/Pr/Ca sublattice, (b) motion of all the Mn$^{4+}$ and oxygen atoms and (c) the Jahn-Teller mode, that is solely motion of the oxygen occupying the distorted octahedra around the Mn$^{3+}$ ions. The gray bar represents the estimated average excitation strength of $\eta = 0.16$.}
\label{fig:calc}
\end{figure*}

In an oxygen isotope Raman study an $A_g$ symmetry optical phonon in the frequency range of 2-2.5 THz observed in $R_{1-x}A_x$MnO$_3$ has been assigned to the motion of the $R/A$ cations.\cite{Amelitchev2001} Consequently, the observed oscillations in time resolved studies in this frequency range have been ascribed to this motion.\cite{Lim2005,Beaud2009,Matsuzaki2009} However, our recent detailed optical pump-probe data indicate that the phonon must instead be connected with the motion of the Mn$^{4+}$O$_6$ sites.\cite{Caviezel2012} In the following we will apply a simple model of displacive excitation to clarify this assignment. The main atomic motions involved when heating from the ordered state at low temperatures to the room temperature structure involve the translational motions of the Mn$^{4+}$O$_6$ sites and the $R/A$ ions, both along the $x$ direction.\cite{Rodriguez2005} In a first step we compare the isolated influence of both of these motions with respect to the measured diffracted x-ray intensity. In a second step we will refine this model in order to describe the measured transients more accurately.

\subsection{Simple displacive excitation model}
Here, we assume a simple displacive mechanism involving the motion along a single structural coordinate.\cite{Zeiger1992} For sufficiently small displacements the normalized diffracted x-ray transients of the superlattice peaks for times $t > 0$ can be approximated by
\begin{equation}
\frac{I\left(t\right)}{I_{0}} \approx 1 + A_{\mathrm{disp}} \left[ 1 -\: \cos (2\pi ft)\:e^{-\frac{t}{\tau}} \right]
\label{eqn:fit1}
\end{equation}
where $A_{\mathrm{disp}}$ is the reduction in diffraction intensity due to the total displacement after approximately 1.5 ps, $f$ the phonon frequency, and $\tau$ its lifetime. Equation \ref{eqn:fit1} is valid only in the limit of small excitation fluences where the diffraction intensity changes nearly linear with the induced displacement. The use of this approximation holds the advantage that the phonon amplitudes and frequencies can be extracted without prior assumption of the underlying atomic motion. In a rigorous treatment the structure factor calculations as discussed in the following must be included. Since we study the response of the same coherent optical phonon experienced by different reflections we use Eqn. \ref{eqn:fit1} to fit the three transients simultaneously to determine its frequency and lifetime more accurately. The temporal resolution of the experiment is taken into account by displacement with a Gaussian of 200 fs (FWHM). The resulting fitted curves are plotted in Fig. \ref{fig:TimeTraces} as solid blue lines. The fit yields $f$ = 2.51(4) THz and $\tau$ = 0.55(6) ps. The values obtained for $A_{\mathrm{disp}}$ are summarized in Table \ref{tab:calc}.

In order to test several scenarios for the phonon assignment we now perform structure factor intensity calculations. Due to the weak diffraction peaks measured a kinematic approximation is justified and results in
\begin{equation}
	I^{hkl}~\propto~\left| F^{hkl}\right|^2 = \left|\sum_{j} f_j\:e^{i\:\mathbf{G}\cdot\mathbf{r}_j}\right|^2
\end{equation}
with the vector $\mathbf{G} = h\mathbf{b}_1 + k\mathbf{b}_2 + l\mathbf{b}_3$ defined via the reciprocal space vectors $\mathbf{b}_1$, $\mathbf{b}_2$ and $\mathbf{b}_3$ and the real space vectors $\mathbf{r}_j = x\mathbf{a}_1 + y\mathbf{a}_2 + z\mathbf{a}_3$ composed of the three lattice vectors $\mathbf{a}_1$, $\mathbf{a}_2$ and $\mathbf{a}_3$. The atomic form factors $f_j$ are based on the tabulated Cromer-Mann coefficients\cite{Paufler2007,Hovestreydt1983} and the real and imaginary part of the dispersion correction.\cite{Henke1993} The calculations take into account the differences between Mn$^{3+}$ and Mn$^{4+}$ ions. Using the x-ray and laser attenuation lengths and accounting for the excitation gradient in the sample we estimate an average excitation fluence in the probed volume of $F_{\mathrm{avg}} \approx 1.6 ~\mathrm{mJ/cm}^2$. For the different peaks we now calculate the changes in diffraction intensity from
\begin{equation}
	\frac{I^{hkl}}{I^{hkl}_0}  =  \frac{\left| \: \sum_j \: f_j \: e^{i\:\mathbf{G} \cdot \left(\mathbf{r}^0_j + \eta \Delta \mathbf{r}_j \right)}    \right|^2}{\left| \: \sum_j \: f_j \: e^{i\:\mathbf{G} \cdot \mathbf{r}^0_j} \right|^2}
	\label{eqn:intensity}
\end{equation}
yielding $A_{\mathrm{disp}} = I^{hkl}/I^{hkl}_0-1$ with the excitation strength $\eta = F_{\mathrm{avg}}/F_{\mathrm{th}} \approx 0.16$ ($\eta = 1$ corresponding to the phase transition) and the structural shift $\Delta \mathbf{r}_j = \mathbf{r}^{\mathrm{RT}}_j - \mathbf{r}^0_j$. The superscript RT denotes the atomic positions of the $Pbnm$ structure at room temperature, and the superscript 0 describes the atomic positions of the low temperature CO/OO structure. We use the room temperature structure and the constrained $P2_1/m$ structure at 20 K of La$_{0.5}$Ca$_{0.5}$MnO$_3$\cite{Rodriguez2005} which is expected to yield a good match for the superlattice reflections of our sample. With Eqn. \ref{eqn:intensity} at hand we calculate the changes in diffraction strength assuming that only specific atoms have moved. It is reasonable to assume that the atoms move along the same directions as they do during the phase transition.

Figure \ref{fig:calc} shows three examples of calculated diffraction intensity changes. Figure \ref{fig:calc}(a) shows the outcome when shifting only the $R/A$ sites, that is the La/Pr/Ca ions. Please note that the very weak ($1\:\nicefrac{1}{2}\:0$) reflection should significantly increase upon laser excitation for this motion, which is clearly not observed in the experiment. Figure \ref{fig:calc}(b) takes into account the motion of all all Mn$^{4+}$ and oxygen atoms and Fig. \ref{fig:calc}(c) assumes only motion of the oxygen octahedra surrounding the Mn$^{3+}$ ions releasing the Jahn-Teller distortion to a more symmetric configuration and is accordingly labeled as the Jahn-Teller stretching mode. 
\begin{table}[tbp]
 \begin{ruledtabular}
 \begin{tabular}{cccccc}
 ($h\:k\:l$)									&Exp: $A_{\mathrm{disp}}$& 			La/Pr/Ca  		&	all Mn$^{4+}$+O		&  all 	\\ \hline
 	(2 -$\nicefrac{3}{2}$ 0)	&     	-0.20(1)				&				-0.11					&			 -0.20				&	-0.3	\\  
 ($2\:\nicefrac{1}{2}\:0$)		& 			-0.24(1)				&				-0.19					&	 			-0.12				&	-0.3	\\  
 ($1\:\nicefrac{1}{2}\:0$)		& 			-0.28(2)				&				+0.30					&	 			-0.53	  		& -0.3	\\ 
 \end{tabular}
 \end{ruledtabular}
  \caption{\label{tab:calc} Relative intensity changes of the Bragg reflections obtained from the experiment and the different model calculations which attribute the phonon motion to either the $R/A$ ions, all the Mn$^{4+}$ and O atoms, and to all atoms involved at an excitation strength of $\eta = 0.16$.}
     \end{table} 
 
The calculated relative changes of the Bragg intensity for specific atom movements at an estimated excitation strength of $\eta = 0.16$ for the three measured reflections are summarized in Table \ref{tab:calc}. The opposite sign in the calculated response of the ($1\:\nicefrac{1}{2}\:0$) reflection excludes the previous assignment of the phonon to the isolated motion of the $R/A$ sites.\cite{Beaud2009} The small contribution calculated for an isolated Jahn-Teller motion [see Fig. \ref{fig:calc}(c)] in addition with its opposite sign and the much higher frequencies assigned to these modes\cite{Matsuzaki2009,Wall2009} also eliminates it as source for the measured low frequency phonon. The comparison of the calculated and experimental values for the intensity changes strongly indicates that the phonon includes the motion of the Mn$^{4+}$ atoms and a subset of the surrounding oxygen atoms. At this point, we cannot rule out the possibility that the phonon could include both the motion of Mn$^{4+}$O$_6$ and the $R/A$ atoms. 

This simple displacive model allowed the exclusion of certain atomic motions attributed to the observed low frequency phonon. However, optical pump probe measurements showed that the initial physical processes in laser excited manganites happen on a significantly faster time scale.\cite{Matsuzaki2009,Wall2009} Taking this into account we analyze our data with a refined two-step displacive excitation model in order to draw more precise conclusions concerning the atomic motion involved.

\subsection{Two-step displacive excitation model}
The instant destruction of the CO/OO caused by photo-induced hot carriers triggers the immediate release of the Jahn-Teller distortion at the Mn$^{3+}$ sites.\cite{Beaud2009,Caviezel2012,Matsuzaki2009} The displacement of the Mn$^{4+}$ sites can therefore not be the primary motion. Indeed high frequency coherent oscillations of displacive character assigned to the Jahn-Teller mode have been observed by optical pump-probe experiments performed with very high time resolution with additional oscillations at intermediate frequencies assigned to the rotational modes of the oxygen octahedra.\cite{Matsuzaki2009, Wall2009} Even though the atomic motions of the oxygens associated with the Jahn-Teller mode are small the corresponding change of Bragg intensity can be significant [see Fig. \ref{fig:calc}(c)]. These initial displacements of the oxygen atoms are significantly faster than our experimental time resolution and subsequently launch the secondary displacement exciting the coherent optical phonon in our experiment. Due to the limited time resolution and the large damping of these fast modes we simply account for the high frequency components by introducing an initial fast displacement to our model that transforms Eqn. \ref{eqn:fit1} for times $t > 0$ into
\begin{equation}
\frac{I\left(t\right)}{I_{0}} \approx 1 + A_{\mathrm{fast}} + A_{\mathrm{ph}} \left[ 1 -\: \cos (2\pi ft)\:e^{-\frac{t}{\tau}} \right]
\label{eqn:fit}
\end{equation}
with $A_{\mathrm{fast}}$ being the reduction of the diffraction intensity due to this initial fast displacement and $A_{\mathrm{ph}}$ the phonon amplitude. The fits using Eqn. \ref{eqn:fit} are represented by the solid red lines in Fig. \ref{fig:TimeTraces}. The phonon frequency and lifetime are evaluated to $f$ = 2.50(4) THz and $\tau$ = 0.53(8) ps and the fitting values for $A_{\mathrm{fast}}$ and $A_{\mathrm{ph}}$ are summarized in Table \ref{tab:final}.
\begin{table*}[t!]
 \begin{ruledtabular}
   \begin{tabular}{@{}ccccccccccc@{}}\toprule
  	\vspace{3pt}
 										&	\multicolumn{2}{c}{Experiment}		    			&			 \multicolumn{2}{c}{Scenario 1}			&		\multicolumn{2}{c}{Scenario 2}	  &	\multicolumn{2}{c}{Scenario 3}	&    \multicolumn{2}{c}{Scenario 4}     \\ \hline
 	\vspace{3pt}
 ($h\:k\:l$)			  				& $A_{\mathrm{fast}}$	&	$A_{\mathrm{ph}}$ & $A_{\mathrm{fast}}$	& $A_{\mathrm{ph}}$	& $A_{\mathrm{fast}}$	& $A_{\mathrm{ph}}$	& $A_{\mathrm{fast}}$ & $A_{\mathrm{ph}}$ & $A_{\mathrm{fast}}$ & $A_{\mathrm{ph}}$\\ \hline
 (2 -$\nicefrac{3}{2}$ 0)		& -0.06(3)				&				-0.14(3)		&		   -0.04	        &		   -0.15      	&			-0.05			&					-0.15			&	-0.04 &			-0.26		& -0.05 &	-0.25				\\  
 ($2\:\nicefrac{1}{2}\:0$)			& 0.10(4)					&				-0.34(4)		&		    0.12		      &		   -0.24        &	 		 0.22			&	 				-0.35			&	0.12	&		  -0.42		&  0.22 & -0.52					\\
 ($1\:\nicefrac{1}{2}\:0$)			& 0.12(12)				&				-0.40(10)		&		    0.01		      &	     -0.54        &			-0.08  		&	  			-0.45		  &	0.01  &			-0.31		&	-0.08 &	-0.22					\\
 \end{tabular}
 \end{ruledtabular}
  \caption{\label{tab:final} Comparison of the experimentally determined and calculated parameters $A_{\mathrm{fast}}$ and $A_{\mathrm{ph}}$ for the three measured superlattice peaks (2 -$\nicefrac{3}{2}$ 0), ($2\:\nicefrac{1}{2}\:0$) and ($1\:\nicefrac{1}{2}\:0$). Scenarios 1 and 2 fit the experimental data the best. See text for definitions of the different scenarios.}
 \end{table*} 
In the following we focus on the most likely scenarios of the atomic motion associated with the observed dynamics. Table \ref{tab:final} gives an overview of the measured and calculated magnitudes for the parameters $A_{\mathrm{fast}}$ and $A_{\mathrm{ph}}$ for the following scenarios.

\begin{figure}[t]
	\centering
		\includegraphics[width=0.44\textwidth]{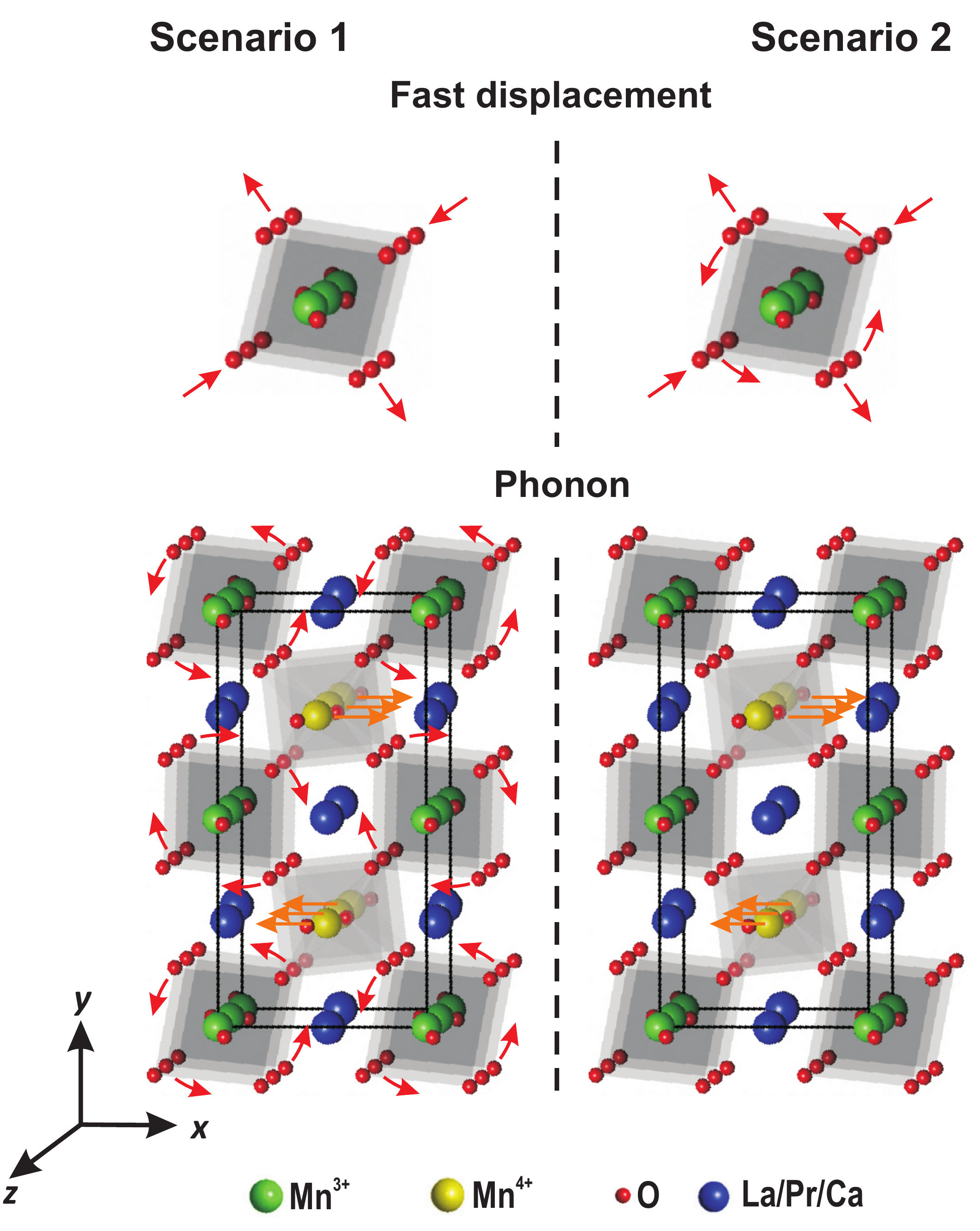}
	\caption{Visualization of the involved atomic motion for the two most likely scenarios.}
	\label{fig:Scenarios}
\end{figure}

Scenario 1, visualized on the left side in Fig. \ref{fig:Scenarios}, is based upon the Jahn-Teller stretching mode as a fast step, that is a partially lifted Jahn-Teller distortion of the Mn$^{3+}$O$_6$ octahedra after the excitation, inducing the motion of the equatorial oxygen atoms near the Mn$^{3+}$ sites. This shift results in a more symmetric configuration of the Mn$^{3+}$O$_6$ octahedra but distorts the octahedral oxygen cage around the Mn$^{4+}$ sites. The latter is compensated by a secondary displacement including the rotation of the equatorial Mn$^{3+}$ oxygens in the $xy$ plane and the motion of the Mn$^{4+}$ atoms with their apex oxygens along $x$. It would be this second displacement that launches the observed coherent phonon.  This is feasible as long as the initial displacement occurs on a time scale significantly faster than the phonon period $1/f = 400$ fs. 

Scenario 2 combines the lifting of the Jahn-Teller distortion and the equatorial rotation of the Mn$^{3+}$ oxygens in the $xy$ plane in the first step. Accordingly, the displacement connected with the coherent phonon is in this case solely the motion of the Mn$^{4+}$ atoms with their apex oxygens in the $x$ direction as shown on the right side in Fig. \ref{fig:Scenarios}.

Additionally, in scenario 3 and 4 the parameter $A_{\mathrm{ph}}$ is evaluated for the assumption that the phonon is ascribed to both the motion of the Mn$^{4+}$O$_6$ and the $R/A$ atoms. That is, scenario 3 is based on  $A_{\mathrm{fast}}$ of scenario 1, scenario 4 on scenario 2. Since none of the predicted phonon amplitudes for scenario 3 and 4 lie within the errors of the experimental data  we can discard them. This supports the assignment of the $R/A$ motion to the slower component observed during the phase transition at high optical fluences.\cite{Beaud2009,Caviezel2012}

The limited time resolution in our experiments and the restricted accuracy of the predicted intensity changes prevents us from unambiguously choosing between scenario 1 and 2. However, optical pump-probe reflectivity measurements with very high time resolution identified oscillations of displacive origin in the range of 6-14 THz (220-470 cm$^{-1}$)\cite{Matsuzaki2009,Wall2009} that correspond to the frequencies of the Jahn-Teller and rotational modes of the oxygen octahedra in good agreement with a Raman study.\cite{Iliev2006} These findings are strongly in favor of scenario 2. Consequently we conclude that the main atomic motion related to the low frequency 1.8-2.5 THz (60-80 cm$^{-1}$) coherent optical phonon observed in laser excited manganites is most probably due to a displacement in the crystal $x$ direction of the Mn$^{4+}$ atoms together with their apex oxygens that bridge neighboring Mn$^{4+}$ sites. 

\section{Conclusions}
In summary, we performed time resolved hard x-ray diffraction measurements on a laser excited LPCMO single crystal. The responses of three superlattice reflections selected with respect to the excited coherent optical $A_g$ phonon associated with the charge and orbital order are investigated. In order to establish the involved atomic motion structure factor calculations are performed for different atomic displacements of the unit cell constituents. Comparison of the calculated intensity changes and the experimental results exclude the assignment of the phonon to the La/Ca/Pr sublattice. These results support the assignment of the observed coherent optical phonon to the Mn$^{4+}$O$_6$ motion as suggested previously. The measured transients strongly favor a two step model. The fast step comprises the release of the Jahn-Teller distortion at the Mn$^{3+}$ sites as well as the rotational motion of the oxygen octahedra. This induces a secondary displacement involving the motion of the Mn$^{4+}$ sites with their apex oxygens triggering the observed low frequency optical phonon.

The main limitations in full structural reconstructions to date are time resolution and flux which restricted our investigations to the brightest superlattice peaks. With the advances of x-ray free electron laser sources it will become feasible not only to study a larger number of reflections more efficiently but also to measure weaker reflections. In addition, with the current efforts to overcome the timing jitter problem\cite{Beye2012} these facilities will permit resolving the fast motion components. In order to fully resolve these dynamics more accurate structure determinations for the exact doping concentration of the analyzed material will also be a crucial ingredient. Finally, our study shows, that despite the low flux at a synchrotron slicing source we are able to clarify the dynamics of a specific phonon in a complex manganite system by investigating a set of carefully chosen reflections.

\begin{acknowledgments}
\vspace{-10pt}
The experiments were performed on the X05LA beam line at the Swiss Light Source, Paul Scherrer Institut, Villigen, Switzerland. This work was supported by the Swiss National Science Foundation (Grant 200021\textunderscore124496) and its National Centers of Competences in Research MUST and MaNEP. We thank the microXAS beamline scientists Daniel Grolimund and Camelia Borca for their support during the experiments. SWC acknowledges support from NSF (DMR-1104484).
\end{acknowledgments}

\bibliography{CaviezelPhononAssignment}

\end{document}